
\raggedbottom
\hoffset=0cm\voffset=0cm
\magnification=\magstep1
\baselineskip=15pt
\def\ref#1{\noindent\item{[#1.]}}

\vskip 9cm
\hfill {\tenrm NUS/HEP/93202}
\par\noindent
\hfill {\tenrm hep-th/9407158}
\vskip 2cm
\centerline {\bf Constraints and Period Relations in Bosonic Strings at
Genus-$g$}
\vskip.7cm
\centerline {C.H. Oh $^{\dag}$ and K. Singh $^{\ddag}$}
\vskip.5cm
{\sl {\centerline {Department of Physics}}}
{\sl {\centerline {Faculty of Science}}}
{\sl {\centerline {Lower Kent Ridge, Singapore 0511}}}
{\sl {\centerline {Republic of Singapore}}}
\vskip 2cm
{\centerline {\bf Abstract}}
\vskip.5cm
\noindent
We examine some of the implications of implementing the usual boundary
conditions on the closed bosonic string in the hamiltonian framework.
Using the KN formalism, it is
shown that at the quantum level, the resulting constraints lead to
relations among the periods of the basis 1-forms. These are compared
with those of Riemanns' which arise from a different consideration.
\vskip 4.5truecm
\centerline {Published in Class. Quantum Grav. {\bf 11} (1994) 503}
\vskip.3cm
\hrule width6cm
\vskip .3truecm
{\sevenrm \dag E-mail: PHYOHCH@NUSVM.NUS.SG
\vskip.1truecm
\ddag E-mail: PHYSINGH@NUSVM.NUS.SG}
\vfil\eject
{\bf \+{1.Introduction.}\hfill \cr}
\par
Formulation of conformal field theories on arbitrary genus
Riemann surfaces has been the focus of intense research in recent
years. Primarily fuelled by a need for a systematic study of
multiloop diagrams in string theories, it has evolved along two
major lines. One approach uses path integrals [1] and the other relies
on operator techniques [2]. Recently, in a series of papers [3,4,5],
Krichever and Novikov (KN) have introduced a new operator formalism
which lends itself to a conceptually simpler and more elegant
treatment of such theories.
Its appealing feature is that it naturally extends the
canonical operator formalism, which has been widely used in studies of
conformal field theories on the sphere, to surfaces of arbitrary genus.
Basically, they introduced global $\lambda$-
differential bases that generalizes the usual Laurent bases which
is global only for the sphere.
Here,
globalization of a theory is obtained by replacing the Laurent basis,
in which fields are usually expanded, to these new bases. The process
of quantization is then carried out in the traditional way in which the
coefficients of the expansions are regarded as operators acting on
some Fock space.
\par
Another important aspect of the KN formalism is that it admits a global
definition of time which in turn allows for a hamiltonian description
of a theory.
Lugo and Russo [6] have given such a formulation of the closed bosonic
string. By using the KN basis, they have defined a
hamiltonian in much the same way as one would do on a sphere and
furnished the corresponding equations of motion.
In this letter, we address some issues that were not taken up in their
paper. Here we examine the boundary conditions and the
resulting implications when the theory is quantized. In particular, we
show that in implementing these conditions at the quantum level,
certain relations among the periods of the basis 1-forms are required.
Apart from these physical requirements, it is further shown that
the same conditions also have a purely topological origin. Indeed,
it turns out that they can be otained directly from the basis
1-forms. Closer examination of these relations, further reveals
strong relationships among the periods which are reminiscent of the
period relations of Riemann. In fact, for a restricted subset of the
periods, these relations appear to be the higher-order analogs of
Riemanns' relations. For $g=1$, in particular, it is shown that the period
relations obtained above reduce to those of Riemanns'.
\+{~}\cr
{\bf \+{2.Some Background and Conventions.} \hfill \cr}
\par
We begin by summarizing some results of Refs.[3,4] pertaining to the
KN-bases. Let $\Sigma$ be a Riemann surface of genus $g$ with two
distinguished points $P_+$ and $P_-$. One can introduce local coordinates
$z_+$ and $z_-$ around these points such that $z_{\pm}(P_{\pm})=0$ .
Now, in accordance with the Riemann-Roch theorem, the basis of meromorphic
functions holomorphic outside $P_+$ and $P_-$ is constructed by specifying
the order of the poles or zeros at these points. Explicitly in the
neighbourhood of $P_{\pm}$ they are given by
$$\eqalignno{A_n(z_{\pm})&=\sum_{m=0}^{\infty}a_{n,m}^{\pm}
z_{\pm}^{\pm n-{g/2}+m}
,~~~~~~~~~~~~~{\vert}n\vert>{g/2},&(2.1a)\cr
A_n(z_{\pm})&=\sum_{m=0}^{\infty}
a_{n,m}^{\pm}z_{\pm}^{\pm n-{g/2}\pm {1/2}-{1/2}+m}
, { \vert n\vert \le {g/2}} ,~~~~~~~~~n\ne {g/2},&(2.1b)\cr
A_n(z_{\pm})&=1 , ~~~~~~~~~~~~~~~~~~~~~~~~~~~~~~~~~~~~~~~ n={g/2},
&(2.1c)\cr}$$
where $a_{n,0}^{+}=1$ and $n$ takes integral values if $g$ is even and
half integral values for odd $g$. Similarly a basis for the space of
differentials in the neighbourhood of $P_{\pm}$ takes the form
$$\eqalignno{\omega_n(z_{\pm})&=\sum_{m=0}^{\infty}
\omega_{n,m}^{\pm}z_{\pm}^{\mp n+{g/2}-1+m}dz_{\pm}
,~~~~~~~~~{\vert}n\vert>{g/2},&(2.2a)\cr
\omega_n(z_{\pm})&=\sum_{m=0}^{\infty}
\omega_{n,m}^{\pm}z_{\pm}^{\mp n+{g/2}\mp {1/2}-{1/2}+m}dz_{\pm}
, { \vert n\vert \le {g/2}} , ~n\ne {g/2},&(2.2b)\cr
\omega_n(z_{\pm})&=\sum_{m=0}^{\infty}
\omega_{n,m}^{\pm}z_{\pm}^{-1+m}dz_{\pm}
 , ~~~~~~~~~~~~~~~~~~ n={g/2} ,&(2.2c)\cr}$$
with the coefficients $\omega_{n,0}^{+}=1, \omega_{g/2,0}^{-}=-1$.
It is worth noting that the differentials are dual to the basis of
meromorphic functions in the following sense,
$${1\over{2\pi i}}\oint_{C_{+}}A_{m}\omega_{n}
= {-}{1\over{2\pi i}}\oint_{C_{-}}A_{m}\omega_{n}=\delta_{mn}
\eqno(2.3)$$
where $C_{+} (C_{-})$ denotes any contour around $P_{+} (P_{-})$ but
not including $P_{-} (P_{+})$. The contours $C_{+}$ and $C_{-}$
can be chosen as the level lines of a single-valued function
$$\tau (P) = 1/2\int _{P_0}^{P}(\omega _{g/2} + \overline {\omega}
_{g/2})\eqno(2.4)$$
such that
$$C_{\tau}=\lbrace Q\in\Sigma
\vert \tau (Q)=\tau,~~~\tau \in {\rm {\bf R}}\rbrace\eqno(2.5)$$
for a fixed $Q_{0}\in\Sigma$, so that as
${\tau\rightarrow\pm \infty}$ they become circles around $P_{\mp}$.
\par
Thus in these bases, any continuously differential function $F(Q)$
and smooth differential $\Omega(Q)$ on $C_{\tau}$ are
given by the following expansions respectively:
$$F(Q)=\sum_{m}A_m(Q)F_m~~~~~~{\rm where}~~~~~~F_m={{1}\over{2\pi i}}
\oint_{C_{\tau}}F(Q')\omega_m(Q')\eqno(2.6)$$
and
$$\Omega(Q)=\sum_{m}\omega_m(Q)\Omega_m~~~~~~{\rm where}~~~~~~
\Omega_m={{1}\over{2\pi i}}\oint_{C_{\tau}}\Omega(Q')A_m(Q').
\eqno(2.7)$$
\par
Next let us briefly recall some results of Ref.[6] that will be useful
in the ensuing discussion. To describe the closed bosonic string it
is convenient to introduce the following parameterization:
$$w(P) = \tau (P) + i \sigma (P). \eqno(2.6)$$
Here $\tau (P)$ is defined in (2.4) and $\sigma (P)$ is given by
$$\sigma (P) = 1/(2i)\int _{P_0}^{P}(\omega _{g/2} - \overline
{\omega} _{g/2}).\eqno(2.7)$$
The phase space for the classical closed bosonic string is defined as
the space of functions $X^{\mu}$ and the differentials $P^{\mu}$, $\mu
= 0,1,2,...D-1$. For a fixed $\tau$ the phase space is characterized by
the poisson bracket
$$\lbrace P^{\mu}(Q),X^{\nu}(Q')\rbrace =-\eta ^{\mu \nu} \Delta
_{\tau}(Q,Q');~~Q,Q'\in C_{\tau}\eqno(2.8)$$
where $\eta ^{\mu \nu}$ is the $D$ dimensional Minkowski metric with
signature (-1,1,...,1) and $\Delta_{\tau}$ is the delta function defined
over $C_{\tau}$,
$$\Delta _{\tau}(Q,Q')=1/(2 \pi i)\sum_{n}\omega _{n}(Q)
A_{n}(Q').\eqno(2.9)$$
The dynamical content of the theory is embodied in the energy momentum
tensor $t$ which is given by the sum of the holomorphic $(T)$ and the
antiholomorphic $(\overline{T})$ parts:
$$T=-1/4(dX + 2\pi P)^2,~~~~
\overline{T}=-1/4(dX - 2\pi P)^2. \eqno(2.10)$$
The hamiltonian, in turn, is obtained from $t$ and is given by
$$\eqalign {H(\tau) &= -1/(2\pi)\oint_{C_{\tau}}(t\vert e_{\sigma})\cr
&= 1/(4 \pi)\oint_{C_{\tau}}(dX^2 + 4\pi^2P^2 \vert
e_{\sigma})}\eqno(2.11)$$
where $e_{\sigma}$ is a meromorphic vector field defined in terms of
vector fields $e_w$ and $e_{\overline{w}}$ that are dual to
$\omega_{g/2}$ and $\overline{\omega}_{g/2}$ :
$$e_{\sigma}=i(e_w - e_{\overline{w}}); \eqno(2.12)$$
$$\eqalignno{(\omega_{g/2}\vert e_w)&= 1 = (\overline {\omega}_{g/2}
\vert e_{\overline{w}})&(2.13a)\cr
(\omega_{g/2}\vert e_{\overline{w})}&= 0 =
(\overline{\omega}_{g/2}\vert e_{w}).&(2.13b)\cr}$$
{}From this hamiltonian, one is led to the equations of motion which, for
$X^{\mu}$, read
$$\partial\overline{\partial}X^{\mu}(Q)=0.\eqno(2.14)$$
For the conjugate momentum, one has
$$P^{\mu}=1/(2\pi)(\partial - \overline{\partial})X^{\mu}.\eqno(2.15)$$
It should be noted that the equation holds for all points on $\Sigma$
except for $P_{\pm}$ and eqn.(2.14), in particular, implies that
$\partial X^{\mu}(\overline {\partial}X^{\mu})$ has to be holomorphic
(antiholomorphic) everywhere except $P_{\pm}$. This in turn suggests
the following expansions:
$$\eqalignno{\partial X^{\mu}(Q)&=(i/\sqrt{2})\sum_{n}\alpha_{n}^{\mu}
\omega_{n}(Q) &(2.16a)\cr
\overline{\partial} X^{\mu}(Q)
&=(i/\sqrt{2})\sum_{n}\overline{\alpha}_{n}^{\mu}
\overline{\omega}_{n}(Q). &(2.16b)\cr}$$
By imposing the boundary conditions
$$\oint_{C_{\tau}}dX^{\mu} =
\oint_{\alpha_{i}}dX^{\mu} =\oint_{\beta_{i}}dX^{\mu} = 0
\eqno(2.17)$$
where $(\alpha_{i},\beta_{i}),~i=1,2,...,g$ is the standard homology
basis, one can integrate $dX^{\mu}$ to give
$$X^{\mu}(Q)=\int_{Q_{0}}^{Q}dX^{\mu}=
x^{\mu}-ip^{\mu}\tau (Q)+(i/\sqrt{2})\sum_{n \ne
g/2}\alpha_{n}^{\mu} B_{n}(Q)+\overline{\alpha}_{n}^{\mu}
\overline{B}_{n}(Q)\eqno(2.18)$$
where $B_{n}(Q)=\int^{Q_{0}}_{Q}\omega_{n}$ and
$x^{\mu}=X^{\mu}(Q_{0})$. It is worth noting that
relations (2.17) essentially ensure the single-valuedness of
$X^{\mu}$. When expansions (2.16a) and (2.16b) are introduced, they
also imply the following relations among the
coefficients $\alpha_{n}^{\mu}$ and $\overline{\alpha}_{n}^{\mu}$:
$$\alpha_{g/2}^{\mu}=\overline{\alpha}_{g/2}^{\mu} =
-p^{\mu}/\sqrt{2} \eqno(2.19)$$
$$\sum_{n}\alpha_{n}^{\mu}a_{n}^{i}+\overline{\alpha}_{n}^{\mu}
\overline{a}_{n}^{i} =0 \eqno(2.20a)$$
$$\sum_{n}\alpha_{n}^{\mu}b_{n}^{i}+\overline{\alpha}_{n}^{\mu}
\overline{b}_{n}^{i} =0 \eqno(2.20b)$$
where
$$a_{n}^{i}=\oint_{\alpha_{i}}\omega_{n};~~~~
b_{n}^{i}=\oint_{\beta_{i}}\omega_{n}.\eqno(2.21)$$
Incorporating these into (2.18) one has
$$X^{\mu}(Q)=
x^{\mu}-ip^{\mu}\tau (Q)+(i/\sqrt{2})\sum_{\vert n \vert
>g/2}\alpha_{n}^{\mu} \phi_{n}(Q)+\overline{\alpha}_{n}^{\mu}
\overline{\phi}_{n}(Q),\eqno(2.22)$$
where $\phi_{n}(Q)$ are harmonic functions given by
$$\phi_{n}(Q)=\int_{Q_{0}}^{Q}(\omega_{n}- \sum_{j=1}^{g}
(F_{nj}\eta_{j}+{\overline{G}}_{nj} {\overline{\eta}}_{j}))
\eqno(2.23)$$
with $\lbrace \eta_{i}\rbrace$ constituting a basis of holomorphic
differentials with normalization
$$\oint_{\alpha_{i}}\eta_{j}=\delta_{ij},~~~
\oint_{\beta_{i}}\eta_{j}=\Omega_{ij},\eqno(2.24)$$
and
$$\eqalignno{F_{nj} &= (i/2)\sum_{i=1}^{g}((\Omega_{2}^{-1}
\overline{\Omega})_{ji} a_{n}^{i}-(\Omega_{2}^{-1})_{ji}b_{n}^{i})
&(2.25a)\cr
G_{nj} &= (i/2)\sum_{i=1}^{g}((\Omega_{2}^{-1}
\overline{\Omega})_{ji}\overline{a}_{n}^{i}-(\Omega_{2}^{-1})_{ji}
\overline{b}_{n}^{i}) &(2.25b)\cr
\Omega_{2} &\equiv {\rm Im}~\Omega. \cr}$$
\par
In quantizing the theory, the usual prescription is to regard the
coefficients $\alpha_{n}^{\mu}$ and $\overline{\alpha}_{n}^{\mu}$
in the above expansions as operators acting on some Fock space and by
replacing the poisson brackets with the corresponding quantum ones
$(\lbrace~,~\rbrace\to 1/i[~,~])$. For (2.8), we have
$$[P^{\mu}(Q),X^{\nu}(Q')] =-i\eta ^{\mu \nu} \Delta
_{\tau}(Q,Q'),~~~Q,Q'\in C_{\tau}\eqno(2.26)$$
which leads to the following commutation relations for the
coefficients:
$$\eqalignno{[\alpha_{n}^{\mu},\alpha_{m}^{\nu}]
=\gamma_{nm}\eta^{\mu
\nu}, ~~~~~~[\overline{\alpha}_{n}^{\mu},\overline{\alpha}_{m}^{\nu}]
&=\overline{\gamma}_{nm}\eta^{\mu \nu},&(2.27a)\cr
[\alpha_{n}^{\mu},\overline{\alpha}_{m}^{\nu}]=0,~~~~~~~~~~~~~~~~
 [x^{\mu},p^{\nu}]&=
-i\eta^{\mu\nu},&(2.27b)\cr}$$
where
$$\gamma_{nm}=(1/2\pi i)\oint_{C_{\tau}}dA_{n}A_{m}.\eqno(2.28)$$
The corresponding Fock space is generated by
$\alpha_{n}^{\mu},\overline{\alpha} _{n}^{\mu}$ with the vacuum defined
by
$$\eqalignno{\alpha_{n}^{\mu}\vert 0>=\overline{\alpha}_{n}^{\mu}\vert
0> &=0,~~~~~n\ge g/2 &(2.29a)\cr
<0\vert \alpha_{n}^{\mu}=<0\vert\overline{\alpha}_{n}^{\mu}
 &=0,~~~~~n<- g/2~~{\rm or}~~n=g/2.&(2.29b)\cr}$$
Note that it is inconsistent to set $\alpha_{n}^{\mu}\vert 0> =
\overline{\alpha}_{n}^{\mu}\vert 0>=0$ for $n \in {\rm I}\equiv
[-g/2,g/2)$ as this will be incompatible with the commutation relations
(since $\gamma_{nm}\ne 0~n,m\in {\rm I}$).
In a quantum theory it is also necessary to introduce normal ordering
in products of operators. For the above oscillators it is defined as
$$:\alpha_{n}^{\mu}\alpha_{m}^{\nu}:=\cases{
\alpha_{n}^{\mu}\alpha_{m}^{\nu}~~~~~&${\rm if}~~n<-g/2~~{\rm
or}~~m>g/2$\cr
\alpha_{m}^{\nu}\alpha_{n}^{\mu}~~~~~&${\rm if}~~n>g/2~~{\rm
or}~~m<-g/2$\cr}\eqno(2.30)$$
As for $n,m \in {\rm I}$, conditions (2.20a) and (2.20b) induce the
following expression:
$$:\alpha_{n}^{\mu}\alpha_{m}^{\nu}:=\alpha_{n}^{\mu}\alpha_{m}^{\nu}
- \eta^{\mu\nu}\sum_{k<-g/2 \atop l>g/2}\sum_{i,j=1}^{g}
(a^{-1})_{in}(a^{-1})_{jm}(\gamma_{kl}F_{il}F_{jk}+
\overline{\gamma}_{kl}G_{il}G_{jk}).\eqno(2.31)$$
With these definitions the hamiltonian is given by
$$H(\tau)=1/2\sum_{n,m}(l_{nm}(\tau):\alpha_{n}^{\mu}\alpha_{m}^{\nu}:
+~\overline{l}_{nm}(\tau):\overline{\alpha}_{n}^{\mu}
\overline{\alpha}_{m}^{\nu}:)\eta_{\mu \nu}\eqno(2.32)$$
where
$$l_{nm}(\tau)=(1/2\pi i)\oint_{C_{\tau}}(\omega_{n}\vert
e_{w})\omega_{m}=l_{mn}(\tau).\eqno(2.33)$$
\+{~}\cr
{\bf \+{3.Constraints.}\hfill \cr}
\par
Having briefly reviewed the formulation of the bosonic string, we will
now study some of the implications of the underlying assumptions.
Note that conditions (2.17), which ensure the
single-valuedness of $X^{\mu}$, are inherently topological as they arise
as a result of $\Sigma$ being non-simply connected. When the appropriate
KN expansions are used, they lead to constraints among the coefficients
$\alpha_{n}^{\mu}$ and $\overline{\alpha}_{n}^{\mu}$. With these
coefficients being regarded as operators in a quantized theory, it
becomes necessary to ask whether these constraints hold as
operator equations. For this to be implemented consistently, one
must require that the constraints commute among themselves.
Moreover, these constraints should not depend on time and must
therefore also commute with the hamiltonian.
\par
Before proceeding further, let us denote the constraints as follows:
$$\eqalignno{\psi^{\mu}
&\equiv \alpha_{g/2}^{\mu}+p^{\mu}/\sqrt{2}=0 ,~~~~~~~
\overline{\psi}^{\mu}\equiv\overline{\alpha}_{g/2}^{\mu}
+p^{\mu}/\sqrt{2}=0&(3.1a)\cr
\Phi_{i}^{\mu}&\equiv\sum_{n}\alpha_{n}^{\mu}a_{n}^{i}
+\overline{\alpha}_{n}^{\mu}\overline{a}_{n}^{i}=0,~~~
\Psi_{i}^{\mu}\equiv\sum_{n}\alpha_{n}^{\mu}b_{n}^{i}
+\overline{\alpha}_{n}^{\mu}\overline{b}_{n}^{i}=0.&(3.1b)\cr}$$
Now, with the constraints being regarded as operator equations, we must
require that they commute with all $\alpha_{n}^{\mu}$,
$\overline{\alpha}_{n}^{\mu}$ for consistency.
For $\psi^{\mu}$ and $\overline{\psi}^{\mu}$,
the commutators are trivially satisfied as
$\gamma_{m~g/2}=\gamma_{g/2~m}=0 $ for every $m$ while for
$\Phi_{i}^{\mu}$ and $\Psi_{i}^{\mu}$ we obtain
$$\eqalignno{[\alpha_{n}^{\mu},\Phi_{i}^{\nu}] &=
\eta^{\mu\nu}\sum_{m}\gamma_{nm} a_{m}^{i},~~~~~
[\overline{\alpha}_{n}^{\mu},\overline{\Phi}_{i}^{\nu}]
=\eta^{\mu\nu}\sum_{m}\overline{\gamma}_{nm}
\overline{a}_{m}^{i},&(3.2a)\cr
[\alpha_{n}^{\mu},\Psi_{i}^{\nu}] &=
\eta^{\mu\nu}\sum_{m}\gamma_{nm} b_{m}^{i},~~~~~
[\overline{\alpha}_{n}^{\mu},\overline{\Psi}_{i}^{\nu}]
=\eta^{\mu\nu}\sum_{m}\overline{\gamma}_{nm}
\overline{b}_{m}^{i},&(3.2b)\cr}$$
which for consistency imply
$$\eqalignno{\sum_{m}\gamma_{nm}a_{m}^{i}&=
\sum_{m}\overline{\gamma}_{nm}\overline{a}_{m}^{i}=0&(3.3a)\cr
\sum_{m}\gamma_{nm}b_{m}^{i} &=
\sum_{m}\overline{\gamma}_{nm}\overline{b}_{m}^{i}=0. &(3.3b)\cr}$$
These conditions are sufficient in ensuring that the
commutators between the the constraints vanish:
$$\eqalignno{[\Phi_{i}^{\mu},\Phi_{j}^{\nu}] &= \eta^{\mu\nu}\sum_{m,n}
(\gamma_{nm}a_{n}^{i}a_{m}^{j}+\overline{\gamma}_{nm}
\overline{a}_{n}^{i}\overline{a}_{m}^{j}), &(3.4a)\cr
[\Phi_{i}^{\mu},\Psi_{j}^{\nu}] &= \eta^{\mu\nu}\sum_{m,n}
(\gamma_{nm}a_{n}^{i}b_{m}^{j}+\overline{\gamma}_{nm}
\overline{a}_{n}^{i}\overline{b}_{m}^{j}), &(3.4b)\cr
[\Psi_{i}^{\mu},\Psi_{j}^{\nu}] &= \eta^{\mu\nu}\sum_{m,n}
(\gamma_{nm}b_{n}^{i}b_{m}^{j}+\overline{\gamma}_{nm}
\overline{b}_{n}^{i}\overline{b}_{m}^{j}). &(3.4c)\cr}$$
Note that $\psi^{\mu}$ and $\overline{\psi}^{\mu}$ commute trivially with
the rest. Likewise, the commutators with the hamiltonian:
$$\eqalignno{[H(\tau),\psi^{\mu}] &= [H(\tau),\psi^{\mu}] = 0,
&(3.5a)\cr
[H(\tau),\Phi_{i}^{\mu}] &= \sum_{n,m,k}(\gamma_{nk}a_{k}^{i}l_{nm}
\alpha_{m}^{\mu}+\overline{\gamma}_{nk}\overline{a}_{k}^{i}
\overline{l}_{nm}\overline{\alpha}_{m}^{\mu}),&(3.5b) \cr
[H(\tau),\Psi_{i}^{\mu}] &= \sum_{n,m,k}(\gamma_{nk}b_{k}^{i}l_{nm}
\alpha_{m}^{\mu}+\overline{\gamma}_{nk}\overline{b}_{k}^{i}
\overline{l}_{nm}\overline{\alpha}_{m}^{\mu})&(3.5c) \cr}$$
also vanish under these conditions.
\par
It is important to note that while eqns.(3.3a) and (3.3b) have emerged as
consistency requirement in the
process of quantization, they are also implied by the KN bases. Indeed,
by setting $\Omega (Q) = dA_n(Q)$ in (2.7) and using (2.28), we have
$$dA_n(Q)=\sum_{m}\gamma_{nm}\omega_m(Q)\eqno(3.6)$$
which essentially expresses an exact form in terms of the basis
1-forms. If we now integrate this relation over a cycle, say
$\alpha_i$, then we find that the integral over $dA_n(Q)$ is zero since
all the periods of an exact 1-form are necessarily zero. The right
hand side, on the other hand, reduces to $\sum_{m}\gamma_{nm}a_{m}^{i}$
which immediately leads us to (3.3a).
Similarly (3.3b) is obtained by integrating (3.6) over the cycle
$\beta_i$.
\par
Another point to be noted is
that the number of terms in (3.3a) and (3.3b) for each $n$ and $i$
are not infinite as it appears but finite because of the following `locality'
conditions:
$$\gamma_{nm}=0 ~~~{\rm for}~~~\cases {~~~~-g&$>n+m>g
,~~~~~~\vert n\vert >g/2,~~\vert m\vert >g/2$ \cr
-g-1&$>n+m>g,~~~{\rm if~either~}\vert n\vert \leq g/2~ {\rm or}
{}~\vert m\vert \leq g/2.$\cr }\eqno(3.7)$$
As a result of these restrictions, we find that for each
$n \in {\rm I}\equiv [-g/2,g/2)$
$$\sum_{m=-g-n-1}^{g-n}
\gamma_{nm}a_{m}^{i}=\sum_{m=-g-n-1}^{\scriptstyle g-n}
\gamma_{nm}b_{m}^{i}=0~~~~~(i=1,2,\ldots g).\eqno(3.8a)$$
By noting that $\gamma_{n~g/2}=0=\gamma_{n~n}$, the above equations
constitute a system of $2g$ equations in $2g$ terms which we write
explicitly as
$$\pmatrix {a_{-g-n-1}^{1} &a_{-g-n}^{1}
&\ldots&a_{g-n}^{1}\cr
\vdots &\vdots &\ddots &\vdots\cr
a_{-g-n-1}^{g}&a_{-g-n}^{g}&\ldots&a_{g-n}^{g}\cr
b_{-g-n-1}^{1}&b_{-g-n}^{1}&\ldots&b_{g-n}^{1}\cr
\vdots&\vdots&\ddots&\vdots\cr
b_{-g-n-1}^{g}&b_{-g-n}^{g}&\ldots&b_{g-n}^{g}\cr}
\pmatrix {\gamma_{n~-g-n-1}\cr
\gamma_{n~-g-n}\cr
\vdots\cr
\vdots\cr
\vdots\cr
\gamma_{n~g-n}\cr}=0.\eqno(3.8b)$$
Note that the elements $\lbrace a_{n}^{i},b_{n}^{i},a_{g/2}^{i},
b_{g/2}^{i}\rbrace_{i=1,2,...,g}$ have been removed since the
$\gamma$'s multiplied to these terms are zero. For $n\not\in {\rm I}$
the summation over $m$ differs for different values of $n$. When
$-3g/2-1\leq n \leq -g/2-1$, $m$ runs over the interval $[-g-n-1,g-n]$ while
for $g/2+1\leq n < -3g/2-1$ it extends over $[-g-n,g-n]$. In both of
these cases we have $2g+1$ terms in $2g$ equations.
\footnote\dag{Note that for $-3g/2-1\leq n \leq -g/2-1$,
$~m=g/2$ lies in the interval $[-g-n-1,g-n]$ so
that the corresponding $a$ and $b$ terms are excluded.}
\par
Clearly, for the set of equations (3.8) to be consistent, one must
require that
$${\rm det}~\pmatrix {a_{-g-n-1}^{1} &a_{-g-n}^{1}
&\ldots&a_{g-n}^{1}\cr
\vdots &\vdots &\ddots &\vdots\cr
a_{-g-n-1}^{g}&a_{-g-n}^{g}&\ldots&a_{g-n}^{g}\cr
b_{-g-n-1}^{1}&b_{-g-n}^{1}&\ldots&b_{g-n}^{1}\cr
\vdots&\vdots&\ddots&\vdots\cr
b_{-g-n-1}^{g}&b_{-g-n}^{g}&\ldots&b_{g-n}^{g}\cr}=0,\eqno(3.9)$$
since, otherwise, the above equations would imply that all the
$\gamma$'s are zero. This is required for all $n \in {\rm I}$ which means
that there are $g$ determinants that must vanish and these lead to
relations among the the periods which are independent of the
$\gamma$'s.
\par
At this juncture it is worth noting that there are other relations
among the periods, namely the period relations of Riemann [7], which have a
different origin from the relations above. Indeed, these
relations are obtained by considering the
{\sl normal form} $\cal M$ of
$\Sigma$ which is obtained by `cutting' $\Sigma$ along the cycles
$\alpha_i$ and $\beta_i$. For a surface of genus $g$, this yields a
$4g$-sided polygon which is simply-connected. If $\theta$ is a
differential of the first or second kind
\footnote\ddag{A differential is of the first kind if it is
analytic everwhere. It is of the second kind if all the residues of its
poles are zero and is of the third kind if it has non-zero residues.}
then we can write $\theta = df$ on $\cal M$. With $f$ being
single-valued on $\cal M$ we have for an arbitrary differential
$\tau$ [7],
$$\oint_{\delta{\cal M}}f\tau =
\sum_{i=1}^{g}[\int_{\alpha_i}\theta \int_{\beta_i}\tau
-\int_{\alpha_i}\tau \int_{\beta_i}\theta] = 2\pi i~({\rm sum~ of~
residues~ of~}(f\tau)).\eqno(3.10)$$
which is the period relation between the differentials $\theta$ and
$\tau$.
\par
For the basis 1-forms $\lbrace \omega_n \rbrace$, with poles at
$P_{\pm}$ we have
$${\cal C}_{nm}\equiv
\sum_{i=1}^{g}(a_{n}^{i}b_{m}^{i}-b_{n}^{i}a_{m}^{i})
=2\pi i~({\rm res}_{P_{+}}
(f_{n}\omega_{m}) + {\rm res}_{P_{-}}(f_{n}\omega_{m}))
{}~~~~n \ne g/2,\eqno(3.11)$$
where $f_n=\int \omega_n$. In the above expression, we have excluded
$n=g/2~$ since the
corresponding differential $\omega_{g/2}$ is a differential of the
third kind and cannot therefore be expressed as an exact form. This is
the only differential of the third kind in the basis. The 1-forms
corresponding to $n\in {\rm I}$ are of the first kind while the rest
are all of the second kind.
\+{~}\cr
{\bf \+{4.Further Analysis.}\hfill \cr}
\par
It is important to note that while the period relations are quadratic
in the periods, relations (3.3a) are linear.
This naturally raises the
question as to whether the above relations imply anything more than
those of Riemann's or, for that matter, whether the two are compatible.
To this end, let us see what these relations imply insofar as (3.3a) are
concerned. To make the analysis more transparent, it is instructive to
consider the $g=1$ case. Besides simplifying the analysis, one also
has the advantage of working with an explicit form for the KN bases.
Indeed, the basis for $\lbrace A_{n}\rbrace$ is given in terms of
of the well studied elliptic functions [8,9]:
$$\eqalignno{A_n(z) &=
{{\sigma^{n-1/2}(z-z_{0})\sigma(z+2nz_{0})}
\over{\sigma^{n+1/2}(z+z_{0})}}{{\sigma^{n+1/2}(2z_{0})}
\over{\sigma((2n+1)z_{0})}},~~~~~n\ne -1/2, &(4.1a)\cr
A_{-1/2}(z) &=
{{\sigma^{2}(z)}
\over{\sigma(z+z_{0})\sigma(z-z_{0})}}{{\sigma(2z_{0})}
\over{\sigma^2(z_{0})}} &(4.1b)\cr}$$
where $\sigma(z)$ is the Wierstrass sigma-function. It should be noted
that the poles are at the points $z=\pm z_{0}$.
Its dual basis
of one-forms $\omega_n$  satisfying (2.3) can be written as
$$\eqalignno{\omega_{n} &= A_{-n}(z)dz,~~~~~~n\ne 1/2 &(4.2a)\cr
\omega_{1/2} &= (A_{1/2}(z)-4\zeta(z_0)+2\zeta(2z_0))dz &(4.2b)\cr}$$
where $\zeta(z)$ is the Wierstrass zeta-function.
In the above bases, one can compute the $\gamma$'s explicitly
using (2.28) and these are given by [8]
$$\eqalignno{\gamma_{nm}(z_0) &= 0,~~~~\vert n+m \vert > 1,~~~(n,m)\ne
(-{1\over 2},-{3\over 2}), &(4.3a)\cr
\gamma_{n,1-n}(z_0) &= n-{1\over 2}, &(4.3b)\cr
\gamma_{n,-n}(z_0)
&= (n+{1\over 2}){\cal D}(2z_0(n+{1\over 2}))+(n-{1\over 2})
{\cal D}(2z_0({1\over 2}-n)), &(4.3c)\cr
\gamma_{n,-1-n}(z_0) &= (n+{1\over 2})[\zeta '(2z_0(n+{1\over 2}))-\zeta
'(2z_0)], &(4.3d)\cr
\gamma_{-1/2.-3/2}(z_0) &= -{{\sigma(4z_0)}\over
{\sigma^{4}(2z_0)}} &(4.3e)\cr}$$
where ${\cal D}(2z_0m)\equiv \zeta (2z_{0}m)-m\zeta (2z_0)$.
\par
Now the equations (3.3a) corresponding to different values of $n$ read
as
\footnote\S {Here we list only the equations with the
$\alpha$-cycles. It should be noted that these equations are paired
with a similar equations in which the $a$-terms are replaced by
the $b$-terms.}
$$\eqalignno{
n=-1/2,~~~~~~
&\gamma_{-1/2,-3/2}~a_{-3/2}+\gamma_{-1/2,3/2}~a_{3/2}=0 &(4.4a)
\cr
n=3/2,~~~~~~~
&\gamma_{3/2,-1/2}~a_{-1/2}+\gamma_{3/2,-3/2}~a_{-3/2}
+\gamma_{3/2,-5/2}~a_{-5/2}=0 &(4.4b)\cr
n=-3/2,~~~~~~&\gamma_{-3/2,-1/2}~a_{-1/2}+\gamma_{-3/2,3/2}~a_{3/2}
+\gamma_{-3/2,5/2}~a_{5/2}=0 &(4.4c)\cr
\vert n\vert \ge 5/2,~~~~~~&\gamma_{n,-1-n}~a_{-1-n}
+\gamma_{n,-n}~a_{-n}+\gamma_{n,1-n}~a_{1-n}=0. &(4.4d)\cr}$$
For $g=1$ eqn.(3.9) reduces to
$${\rm det}\pmatrix {a_{3/2} &a_{-3/2}\cr
                     b_{3/2} &b_{-3/2}\cr}=0 \eqno(4.5)$$
which is precisely the condition that ${\cal C}_{3/2,-3/2}$ vanishes.
Before evaluating this, it is interesting to note that
eqns. (4.4b) and (4.4c) also lead to a
similar vanishing determinant. To see this, we elliminate the $a_{-1/2}$ term
from (4.4b) and (4.4c)
and by using (4.4a) we obtain
$$\gamma_{-3/2,-1/2}~~\gamma_{-5/2,3/2}~~a_{-5/2}-\gamma_{3/2,-1/2}~~
\gamma_{5/2,-3/2}~~a_{5/2}=0 \eqno(4.6)$$
which together with its $\beta$-cycle counterpart requires that
$${\rm det}\pmatrix {a_{5/2} &a_{-5/2}\cr
                     b_{5/2} &b_{-5/2}\cr}=0. \eqno(4.7)$$
In fact by pairing equations for $n$ and $-n$ in (4.4d), it can be shown
inductively that
$$\gamma_{-3/2,-1/2}(\prod_{m=5/2}^{n}\gamma_{-m,m-1})~a_{-n}
+
(-1)^{n-3/2}\gamma_{3/2,-1/2}(\prod_{m=5/2}^{n}\gamma_{m,1-m})~a_{n}=0
\eqno(4.8)$$
for $n=5/2,7/2....$ This means that one must, in general, have
$${\rm det}\pmatrix {a_{n} &a_{-n}\cr
                     b_{n} &b_{-n}\cr}=0 \eqno(4.9)$$
which in turn demands that ${\cal C}_{n,-n}$ vanishes.
\par
Now let us calculate ${\cal C}_{n,-n}$ for $\vert n\vert \ge 3/2$. To
do this, we need an explicit representation of $f_n$. Since $f_n$
is only required in the vicinity of $P_{\pm}$, we have
$$f_n(z_{\pm})=\int A_{-n}(z_{\pm})dz_{\pm}
=\sum_{k=0}^{\infty}{{a_{-n,k}^{\pm}
}\over
{k\mp n+1/2}}z_{\pm}^{k\mp n+1/2},\eqno(4.10)$$
where the coefficients $a_{n,k}^{\pm}$ can be obtained from the
following integral:
$$a_{n,k}^{\pm} = \oint_{\pm z_0}{{dz}\over {2 \pi i}} (z\mp
z_0)^{\mp n-k-1/2} A_n (z). \eqno(4.11)$$
Then by using the above form for $f_n$, the residues in (3.11) can
be computed and this gives
$${\cal C}_{n,m}=2\pi i~\lbrace \sum_{l=0}^{n+m-1}
{{a_{-n,n+m-1-l}^{+}a_{-m,l}^{+}}\over {m-l-1/2}}+\sum_{l=0}^{-n-m-1}
{{a_{-n,-n-m-1-l}^{-}a_{-m,l}^{-}}\over {-m-l-1/2}}\rbrace.
\eqno(4.12)$$
Note that the first sum is zero for $n+m<1$ while the second is
zero for $n+m>1$. From this we can surmise that ${\cal C}_{n,m}=0$
for $m=-n$ which means that both (4.5) and (4.9) are compatible
with the period relations. This, however, does not
conclusively prove that all the equations are consistent. Indeed,
relation (4.8) only replaces one of the two paired equations
(corresponding to $n$ and $-n$ in (4.4d)). For full consisitency,
one must also show compatibility for the remaining equations.
To this end consider (4.4d) with $n$
replaced by $-n$:
$$\gamma_{-n,1+n}~a_{1+n}+\gamma_{-n,n}~a_{n}+
\gamma_{-n,n-1}~a_{n-1}=0\eqno(4.13)$$
with the corresponding $\beta$-cycle equation:
$$\gamma_{-n,1+n}~b_{1+n}+\gamma_{-n,n}~b_{n}+
\gamma_{-n,n-1}~b_{n-1}=0.\eqno(4.14)$$
Then by multiplying (4.13) by $b_{m}$ and (4.14) by $a_{m}$ and
subtracting, we have
$$\gamma_{-n,1+n}~{\cal C}_{1+n,m}+\gamma_{-n,n}~{\cal C}_{n,m}+
\gamma_{-n,n-1}~{\cal C}_{n-1,m}=0\eqno(4.15)$$
which should hold for all $n$ and $m$. It is not very easy to
verify this for generic values of $n$ and $m$, particularly when
$\vert n+m \vert \gg 1$, since this entails computing a large number
of coefficients $a_{n,k}^{\pm}$ in the sum of eqn.(4.12). However, when
$\vert n+m \vert$ is small, the task becomes tractable. For instance when
$n+m=0$, we have
$$\eqalignno{{\cal C}_{1+n,-n} &= 2\pi
i~{{a_{-n,0}^{+}a_{n,0}^{+}}\over {-n-1/2}}=-{{2\pi i}\over {n+1/2}}
&(4.16a) \cr
{\cal C}_{n-1,-n} &= 2\pi
i~{{a_{1-n,0}^{-}a_{n,0}^{-}}\over {n-1/2}}={{2\pi i}\over {n-1/2}}
{}~{{\sigma^2(2z_0)\sigma^2((2n-1)z_0)}\over
{\sigma((2n+1)z_0)\sigma((2n-3)z_0)}}
.&(4.16b) \cr}$$
Then by using the identity (see Ref.[9])
$$\wp(u)-\wp(v)=
-{{\sigma(u+v)\sigma(u-v)}\over{\sigma^{2}(u)\sigma^2(v)}}
\eqno(4.17)$$
where $\wp (u)=-\zeta '(u)$ is the Wierstrass $\wp$ function, one
can show that (4.15) with $m=-n$ holds for all $\vert n \vert \ge
3/2$. Similarly we have also verified the equation for $n+m=\pm 1$.
\+{~}\cr
{\bf \+{5.Concluding Remarks.}\hfill \cr}
\par
In the analysis for $g=1$ above, we have shown that for (3.9) at least,
there is no inconsistency with the corresponding period relation. In
fact the condition coincides exactly with the period relation. The
situation for $g>1$ is not quite the same. Here we have $g$ vanishing
determinants, each resulting in an equation of order $2g$ in the
periods. The period relations of Riemann on the other hand are always
quadratic in the periods irrespective of $g$. For the $2g$ periods
appearing in each
determinant, we have $g(2g-1)$ period relations that must be
compatible. For example, in the $g=2$ case the two vanishing
determinants are
$${\rm det}~\pmatrix {
a_{-2}^{1} &a_{0}^{1} &a_{2}^{1} &a_{3}^{1}\cr
a_{-2}^{2} &a_{0}^{2} &a_{2}^{2} &a_{3}^{2}\cr
b_{-2}^{1} &b_{0}^{1} &b_{2}^{1} &b_{3}^{1}\cr
b_{-2}^{2} &b_{0}^{2} &b_{2}^{2} &b_{3}^{2}\cr}~~~~~~~~~{\rm and}
{}~~~~~~~~~
{\rm det}~\pmatrix {
a_{-3}^{1} &a_{-2}^{1} &a_{-1}^{1} &a_{2}^{1}\cr
a_{-3}^{2} &a_{-2}^{2} &a_{-1}^{2} &a_{2}^{2}\cr
b_{-3}^{1} &b_{-2}^{1} &b_{-1}^{1} &b_{2}^{1}\cr
b_{-3}^{2} &b_{-2}^{2} &b_{-1}^{2} &b_{2}^{2}\cr}$$
which are quartic in the periods.
The first determinant should be compared with six period relations
associated with the 1-forms $\lbrace
\omega_{-2}$, $\omega_{0}$, $\omega_{2}$, $\omega_{3}\rbrace$ i.e.
$\lbrace {\cal C}_{-2,0}$, ${\cal C}_{-2,2}$, ${\cal C}_{-2,3}$,
${\cal C}_{0,2}$, ${\cal C}_{0,3}$, ${\cal C}_{-1,2}\rbrace$
while the second determinant shoud be checked with those among
$\lbrace \omega_{-3}$, $\omega_{-2}$, $\omega_{-1}$,
$\omega_{2}\rbrace$.
We would also like to remark that conditions (3.3a) and (3.3b) appear
to be very strong in the following sense. Apart from being linear,
they also relate the $a$ and the $b$ periods separately. In the $g=1$ case
for instance, all the $a$-periods can be expressed in terms of the
$a_{-1/2}$ and $a_{3/2}$ (or $a_{3/2}$) through the equations
(4.4a-4.4d) and (4.8). The $b$-periods are also similarly related. The
period relations of Riemann on the other hand, relate the $a$ and $b$
periods of two 1-forms quadratically.
\par
It is also interesting to note that the conditions (3.3a) and (3.3b)
also arise under more general boundary conditions. If we allow
$X^{\mu}$ to be multi-valued then conditions (2.17) are replaced by
[10]
$$\oint_{C_{\tau}}dX^{\mu} = 2\pi l^{\mu},~~~
\oint_{\alpha_{i}}dX^{\mu} =2\pi n^{\mu}_{i},~~~
\oint_{\beta_{i}}dX^{\mu} = 2\pi m^{\mu}_{i} $$
where $l^{\mu},m_{i}^{\mu},n_{i}^{\mu}\in {\bf {\rm Z}}$. It is easy to
see that the commutators as in (3.2a) and (3.2b) will also reduce the
corresponding constraints to (3.3a) and (3.3b).
\par
Finally to summarize briefly, we have shown that in fulfilling the boundary
conditions in the closed bosonic string at genus $g$, certain
constraints on the coefficients of the fields and the periods are
implied. At the quantum level, the implementation of these constraints
as operator equations lead to linear relations among the periods of the
basis 1-forms. We have further shown that these relations are also
intrinsic to the the basis 1-forms. Some of these relations imply
relationships among the periods that appear to be the higher-order
analogs of Riemann's period relations. In the $g=1$ case, the
relations turn out to be precisely those of Riemanns'.
\vfil\eject
{\bf REFERENCES}
\ref 1 Polyakov A.M. (1981) Phys. Lett. {\bf B103}  207, 211.
\ref 2 Ishibashi N., Matsuo Y. and Ooguri H. (1987) Mod. Phys. Lett {\bf A2}
119; Vafa C. (1987) Phys. Lett. {\bf B190} 47.
\ref 3 Krichever I.M. and Novikov S.P. (1987) Funct. Anal. Pril. 21
No.2 46.
\ref 4 Krichever I.M. and Novikov S.P. (1987) Funct. Anal. Pril. 21
No.4 47.
\ref 5 Krichever I.M. and Novikov S.P. (1989) Funct. Anal. Pril. 23
No.1 24.
\ref 6 Lugo A. and Russo J. (1989) Nucl. Phys. {\bf B322} 210.
\ref 7 Cohn H. (1967) {\sl Conformal Mapping on Riemann Surfaces},
(McGraw-Hill).
\ref 8 Mezincescu L., Nepomechie R. I. and Zachos C.K. (1989) Nucl.
Phys. {\bf B315} 43.
\ref 9 Chandrasekharan K. ( 1985) {\sl Elliptic Functions}, (Springer Verlag).
\ref {10} Russo J. (1989) Mod. Phys. Lett. {\bf A4} 2349.
\vfil\bye